# Spectrally-resolved dielectric function of amorphous and crystalline GeTe nanoparticle thin films

Ann-Katrin U. Michel,[1] Marilyne Sousa,[2] Maksym Yarema,[3] Olesya Yarema,[3] Vladimir Ovuka,[3] Nolan Lassaline,[1] Vanessa Wood,[3] and David J. Norris[*,1]

[1] Optical Materials Engineering Laboratory, Department of Mechanical and Process Engineering, ETH Zurich, 8092 Zurich, Switzerland.

[2] IBM Research-Zurich, Rüschlikon, Switzerland.

[3] Materials and Device Engineering Group, Department of Information Technology and Electrical Engineering, ETH Zurich, 8092 Zurich, Switzerland.

* E-mail: dnorris@ethz.ch.

Phase-change materials (PCMs), which are well-established in optical and random-access memories, are increasingly studied for emerging topics such as brain-inspired computing and active photonics. These applications take advantage of the pronounced reflectivity and resistivity changes that accompany the structural transition in PCMs from their amorphous to crystalline state. However, PCMs are typically fabricated as thin films via sputtering, which is costly, requires advanced equipment, and limits the sample and device design. Here, we investigate a simpler and more flexible approach for applications in tunable photonics: the use of sub-10 nm colloidal PCM nanoparticles (NPs). We report the optical properties of amorphous and crystalline germanium telluride (GeTe) NP thin films from the infrared to the ultraviolet spectral range. Using spectroscopic ellipsometry with support from cross-sectional scanning electron microscopy, atomic force microscopy, and absorption spectroscopy, we extract refractive indices $n$, extinction coefficients $k$, and band gaps $E_g$ and compare to values known for sputtered GeTe thin films. We find a decrease of $n$ and $k$ and an increase of $E_g$ for NP-based GeTe films, yielding insights into size-dependent property changes for nanoscale PCMs. Furthermore, our results reveal the suitability of GeTe NPs for tunable photonics in the near-infrared and visible spectral range. Finally, we studied sample reproducibility and aging of our NP films. We found that the colloidally-prepared PCM thin films were stable for at least two months stored under nitrogen, further supporting the great promise of these materials in applications.

Phase-change materials (PCMs) can exhibit rapid and reversible transitions between their amorphous (a) and crystalline (c) states. These structural transitions modify their electrical and optical properties dramatically. For example, the resistivity can change by up to six orders of magnitude, an effect that has been employed in electronic[1] and neuromorphic memories.[2,3] The refractive index change, $\Delta n = |n_c - n_a|$, can be up to ~3.6, underlying the successful deployment of these materials in rewriteable optical data storage.[4,5] Moreover, the ability to control the refractive index in PCMs either electrically or optically has motivated further exploration for applications in photonics.[6-11] However, PCMs, which are typically based on chalcogenides (*e.g.* Te) and/or pnictogens (*e.g.* Sb), are commonly deposited as thin films by sputter deposition and patterned by conventional lithography. This results in a costly process, requiring specialized equipment and a high level of technical expertise. If more straightforward and inexpensive approaches to these materials were available, their utility in tunable photonics[9,12,13] could be greatly enhanced.

One possible strategy would be to exploit films of colloidal nanoparticles (NPs) made from a PCM. If such particles could be easily synthesized and deposited on planar or patterned[14] substrates, they could increase access to this material class. To examine this possibility, here we investigate the preparation and characterization of PCM thin films based on colloidal NPs of germanium(II) telluride, GeTe. Specifically, we study particles with diameters below 10 nm. Because of their reduced size, the properties of these NPs can be affected compared to bulk GeTe. For example, localized surface plasmon resonances can arise in crystalline (c-)GeTe NPs.[15] The crystallization temperature, $T_C$, can also change. Data in the literature indicates that when the diameter of amorphous (a-)GeTe NPs is decreased from 8 to 2 nm, $T_C$ increases from 215 to 400°C.[16-18] An increase of $T_C$ was also reported for sputtered PCM films thinner than 10 nm.[19,20] This effect was explained by the bonding characteristics in these materials,[21] but the unique bonding in PCMs is still debated.[22,23] Consequently, an in-depth study of small-size PCM NPs will not only promote the application of these materials but also address fundamental scientific questions.

To move toward a deeper understanding of PCM NPs, we report a detailed study of their optical properties. We collected spectroscopic ellipsometry (SE) data from thin films of GeTe NPs in their amorphous as well as crystalline phases to determine their refractive index $n$, extinction coefficient $k$, and optical band gap $E_g$. Additionally, we investigated the effect of sample age on the optical properties of GeTe NP films. Sample integrity is a crucial aspect for experimental studies, since sample stability over many weeks or months facilitates the study and application of these materials.

The extraction of $n$ and $k$ from thin sputtered films of PCMs[24-26] or NP-based films of indium tin oxide[27] or semiconductor quantum-dots (QDs)[28,29] has been discussed in the literature. Typically, such studies rely on SE as a reliable, well-recognized technique. An alternative approach would be to derive an empirical model and extract the optical properties from reflectance and transmission spectra. However, this would require many fits without any benefit over standard SE.[24] Thus, we chose spectroscopic ellipsometry as our measurement technique.

We note that the QD studies referenced above, as well as the work presented here, involve the direct measurement of a composite $n$ and $k$. No distinction is made between the NPs in the film, the molecules (or ligands) attached to or surrounding the NPs, and any potentially existing pores. This composite (or effective) refractive index is not only easier to access without precise knowledge of all the individual material constituents in the composite film, but is also a more useful quantity for optoelectronic devices or photonic applications, where the NP-based films could be applied as active layers. Accordingly, we chose a sample stack as close as possible to published metamaterial designs involving PCM films.[30,31]

To study the dielectric function (relative permittivity) of thin films composed of PCM NPs, nearly monodisperse a-GeTe NPs were prepared by employing an amide-promoted synthetic approach described previously.[16] Other reported routes lead to more polydisperse,[17] much larger,[32,33] or crystalline GeTe NPs.[34,35] Our particles, which remain amorphous after synthesis, can be easily transformed to their crystalline state by annealing a deposited NP film on a hot plate. Thus, we had access to both structural phases of this model PCM nanomaterial. Furthermore, by working with particles that have a diameter below 10 nm, we can study a regime for which size-dependent-property scaling has been reported, *e.g.* for $T_C$.[16,17,19,20]

Our GeTe NPs had an average size of 5.8 ± 0.6 nm, as can be seen from a typical transmission electron microscopy (TEM) image [Fig. 1(a)] as well as from the particle size distribution shown in the supplementary material. For all further steps below, silicon substrates with 100 nm of native oxide ($SiO_2$) were used. GeTe NPs were spin-coated onto the substrates to form a thin film and capped with an additional sputtered layer of $SiO_2$ (≈ 38 nm). The latter is necessary since GeTe is known to degrade easily upon annealing in ambient atmosphere, potentially leading to chemical segregation of the alloy.[36,37] To study the crystalline phase of GeTe, a subset of the samples was annealed at 300°C, which is above the $T_C$ reported earlier for this material ($T_{C,NP}$ ≈ 225°C).[16] Crystallization was indicated by a color change of the sample surface and confirmed by X-ray diffraction [Fig. 1(b)]. We note that the latter indicates NP fusion upon annealing. A more detailed description of the samples and their fabrication can be found in the supplementary material.

Cross-sectional scanning electron micrographs (SEMs) of the as-deposited [Fig. 1(c)] and annealed [Fig. 1(d)] films reveal the thickness of each deposited layer: 42 nm for the a-GeTe sample $S_{a1}$ and 29 nm for the c-GeTe sample $S_{c1}$ [$t_{exp}$ in Tab. 1]. Even though the continuous NP layer shows a variation in its height profile, which was partially transferred to the topography of the $SiO_2$ cap (cf. supplementary material), no influence on the ellipsometry data was expected, since the elliptical area probed by SE can be as large as (1×8) mm². Thus, height variations as seen in Figs. 1(c) and 1(d) will be averaged, which is reflected in Fig. 2(a). There, the raw data collected with SE is given by $\Psi(E)$ and $\Delta(E)$. These describe the change in polarization of the linearly polarized incident beam, containing *s*- and *p*-polarized light, due to interaction with the sample surface as a function of the energy *E*. $\Psi$ is defined as the amplitude ratio upon reflection of the *s*- and *p*-polarized components, while $\Delta$ represents the polarization phase difference between them. $\Delta$ is very sensitive to tiny changes in the film thickness or material microstructure.[38] For very simple sample structures, the amplitude ratio $\Psi$ is characterized by the refractive index *n*, while $\Delta$ directly determines light absorption described by the extinction coefficient *k*.[38] In Fig. 2(a), $\Psi(E)$ and $\Delta(E)$ collected at three different positions on the c-GeTe sample (translation with automated stage by 1.5 mm in *x* and *y*-direction relative to the center position) are compared. Essentially no differences in $\Psi$ and only minor differences in $\Delta$ are observed.

Additionally, we fabricated multiple samples from the same NP batch to study reproducibility. For a better overview, below we only show data for $\Psi(E)$ collected at 75°, the angle closest to the Brewster angle for the studied samples. At this angle, the measured signal is maximized.[38] While no differences between the data for two a-GeTe samples, $S_{a1}$ and $S_{a2}$, can be identified [Fig. 2(b)], small deviations in $\Psi$ around 1.5 eV are seen for two c-GeTe films, $S_{c1}$ and $S_{c2}$ [Fig. 2(c)]. These small differences in $\Psi$ are likely due to the different film roughness in the c-GeTe NP films (cf. supplementary material).

The measured spectra of the GeTe NP films were fit by an oscillator model using the WVASE® software.[39] The best spectral description with a minimum number of oscillators was achieved by applying six Gaussians and two parameterized semiconductor oscillators. Furthermore, surface roughness was included to generate a better fit in the ultraviolet spectral range. All oscillator parameters were chosen carefully not only to ensure that the fits would avoid local minima, but also to result in layer thicknesses consistent with cross-sectional SEMs [Tab. 1]. A Tauc-Lorentz oscillator with a Drude term, as used for sputtered PCM films,[24] did not fit the measured SE data from our NP thin films (both before and after annealing). This might be caused by the different sample morphology and constituents. While thick sputtered GeTe layers with a smooth surface can be assumed to consist of nearly 100% GeTe, randomly arranged spherical NPs build films with a rougher surface and are composed not only of GeTe NPs, but also of ligands, solvent residues, and pores. The contributions are likely to change upon annealing (cf. supplementary material). This sample architecture includes more constituents with potential interactions and optical transitions than the sputtered films. Thus, the free charge carriers and the onset of optical transitions described by a Drude-type contribution and a Tauc-Lorentz oscillator, respectively, are not sufficient to describe the NP-based film. Combining Tauc-Lorentz with other oscillators is mathematically possible, but is likely to lead to artifacts without any physical meaning.[40] Therefore, we chose the common approach of applying multiple harmonic or Gaussian oscillators to describe the GeTe NP films and the different optical transitions therein.[28,29] While they fit the experimental SE data well, further experimental and numerical studies would be necessary to assign a physical transition to each oscillator.

In Fig. 3(a), the measured data for $\Psi(E)$ from the a-GeTe NP film is compared to the resulting fit. Despite the overestimation of $\Psi$ at $E \approx 1.25$ eV and the small fitting artifacts at 1.2 and 2.2 eV, the experimental data are described well. The fit for the c-GeTe NP film is also in excellent agreement with the measured ellipsometry data [Fig. 3(b)].

The extracted refractive index $n(E)$ and extinction coefficient $k(E)$ of a- and c-GeTe NP films are shown in Figs. 3(c) and (d), respectively. The films show qualitatively similar behavior: while pronounced features can be found for $n$ and $k$ in the near infrared (IR) spectral range (near 1 eV), both quantities are close to 2 and remain spectrally flat in the visible. With increasing energy into the ultraviolet (UV), they show an increase and decrease in $n$ and $k$, respectively. While $n_c$ exhibits a smoother behavior, small sharp features are visible in $n_a$ at about 1.2 and 2.2 eV. Both can be ascribed to the small fitting artifacts discussed above [Fig. 3(a)]. Thus, we exclude these features in $n_a$ from any further discussion. The extinction coefficient, which is associated with absorption, is below 0.5 for $E > 1$ eV for the a-GeTe NP film and for the entire spectral range for the crystalline state. Moreover, $k_a = 0$ between ~1.10 and 1.36 eV, while $k_c = 0$ for $E < 0.78$ eV.

In Figs. 3(e) and (f), we compare our data to the refractive index and extinction coefficient of sputtered GeTe films in their amorphous (A) and crystalline (C) state as reported by Shportko *et al.* (GeTe film thickness: 0.5 µm as-deposited).[24] To distinguish the data from NP and sputtered films, $n$ and $k$ have lower- and upper-case subscripts, respectively. In general, the sputtered film features much higher $n_{A/C}$ and $k_{A/C}$ with a stronger variation over the studied energy range relative to the NP films. We suggest that these differences can be related to the *volume fraction f*.[41] While we can assume that ideal sputtered films have no pores as deposited ($f = 100\%$), an upper limit $f_{max} = 64\%$ for a random packing of spheres is expected.[42] Accordingly, we applied multiple effective medium approaches and made a rough estimate of $f$ being around 60% for the as-deposited NP-based GeTe films (cf. supplementary material). While a reduced volume fraction might explain the differences between sputtered and NP films regarding $n$ and $k$ in the visible, the deviations in the IR and UV spectral range remain unclear. In the UV, sample surface texture has a strong influence.[38] Extending SE measurements to the UV as well as careful determination of the sample surface profile might give further insight. In the IR, effects from molecules (or ligands) surrounding the NPs, free charge carriers, and the band gap $E_g$ of PCMs can be found,[24] and thus complicating the interpretation of the measured $n$ and $k$ in this spectral range. In fact, in Figs. 3(e) and (f), the lowest energy for which $k = 0$ (or has a minimum) is shifted to higher energies for the NPs compared to the reference data on thick sputtered films. This could indicate a shift of the band gap $E_g$ for the NPs relative to the sputtered GeTe.

Since property scaling has been found for GeTe NPs compared to bulk materials,[15-17] a further analysis of the NP band gap, $E_{g,a/c}$, is of high interest. Figure 4(a) shows the absorption spectrum for a colloidally-stable dispersion of a-GeTe NPs (*i.e.* the NPs are physically separated in a non-polar solvent). Figure 4(b) gives the related Tauc plot, which considers the relationship between the absorption coefficient α, the photon energy $E$, and $E_g$: $\alpha E \propto (E - E_g)^r$.[43-45] The exponent $r$ is chosen according to the character of the transition, *e.g.* allowed or forbidden, direct or indirect. Accordingly, the Tauc plot shows $(\alpha E)^{1/r}$ as a function of $E$. Linear interpolation reveals the band gap as the abscissa in this plot. If linearity is found in the data, the assumed transition, expressed by the choice of $r$, can be assumed to be valid. The linear extrapolation in Fig. 4(b) allows for an approximation of the optical band gap $E_{g,a} \approx 1.17$ eV, which is 0.39 eV larger than the reported $E_{g,A} = 0.78$ eV for sputtered GeTe.[24]. We assumed an indirect allowed transition ($r = 2$), since it has been shown to be valid for A-PCMs, such as A-Ge$_2$Sb$_2$Te$_5$ and A-GeTe.[43,44]

To determine $E_g$ for the NP-based films, we again used the Tauc plot plus two additional approaches that are well-established in literature: the heuristic method suggested by Böer,[45] and the evaluation of $k$.[24] For the Tauc plot, we chose $r = 2$ as above, and found $E_{g,a} \approx 1.37$ eV and $E_{g,c} \approx 0.78$ eV. The heuristic approach by Böer defines the band gap as the energy for which the absorption coefficient α reaches $10^4$ cm$^{-1}$.[43,45] We determined $E_{g,a} \approx 1.43$ eV and $E_{g,c} \approx 0.82$ eV. Finally, we evaluated the extinction coefficient spectrum and determined the onset of $k > 0$. The latter has been successfully applied to sputtered PCM thin films.[24] For our NP-based films, we found $E_{g,a} \approx 1.37$ eV and $E_{g,c} \approx 0.78$ eV. In the supplementary material, we discuss all three methods for the NP films in more detail. The averaged estimated values for $E_{g,a}$ and $E_{g,c}$ for the a- and c-GeTe NP films are about 0.61 eV and 0.24 eV larger than the reported $E_{g,A} = 0.78$ eV and $E_{g,C} = 0.55$ eV for sputtered GeTe, respectively.[24] Interestingly, size-dependent scaling of the band gap is well known for QDs (*i.e.* the quantum confinement effect).[46,47] The measured increase in bandgap for colloidal PCM particles and films based on these could provide similar opportunities to explore emerging phenomena at the nanoscale.

For tunable photonics not only the band gap but the refractive index contrast $\Delta n$ and the figure of merit (FOM) $\Delta n/k_A$ are important measures.[48,49] When PCMs are to be combined with plasmonic nanostructures that strongly interact with light, a large $\Delta n$ results in a large spectral shift of the respective plasmonic resonance.[50] Furthermore, a large FOM minimizes the damping of this resonance for a given shift. In Fig. 4(c), we evaluate $\Delta n$ and the FOM at 600 nm as well as the band gap $E_{g,a/A}$ for our NP-based films, the sputtered GeTe films described by Shportko *et al.*,[24] and the PCM Ge$_2$Sb$_2$Te$_5$,[49] which is widely applied in photonics. For the band gap, which indicates the energy for which absorptive losses start, and the FOM, the GeTe NPs are superior compared to the two other PCMs. However, the improved FOM is likely due to the higher volume occupied by air in the NP-based films. In turn, this leads to the relatively low $\Delta n$ for the GeTe NP films. Nevertheless, the radar chart for the GeTe NP-based films [Fig. 4(c)] is promising regarding the recent interest in GeTe for photonics.[52] Moreover, by controlling the volume fraction $f$ in the NP films, one obtains a knob to tune $n$ and $k$, which is not possible for sputtered PCM films. To exploit the GeTe NP films for active photonics or storage applications, reversible phase transitions, *e.g.* by the means of optical pulses would be beneficial.[14]

Finally, we studied the influence of aging of the NP-based films stored at room temperature under nitrogen. Stable optical properties are crucial for further application of these films, *e.g.* in photonic and memory applications. We repeated SE on our NP-based a-GeTe and c-GeTe samples three and eight weeks after film fabrication. Ψ(*E*) did not show any variation over time, which can be seen in Fig. 5. Additionally, Δ(*E*) for aged samples is shown in the supplementary material. Based on the SE spectra, we conclude that the NP-based thin films are stable for at least two months. This could allow for their application in tunable metamaterials,[52] such as color displays, without the disadvantage of changes in their optical properties.

In conclusion, we studied the optical properties of reproducible and stable GeTe NP thin films in their amorphous and crystalline state. We determined the effective refractive indices, extinction coefficients, and band gaps and found clear shifts relative to sputtered films (smaller *n* and *k*, larger $E_g$). This could potentially be ascribed to the relatively smaller volume fraction of the NP-based PCM layer and to a confinement effect due to the small size of the NPs, respectively. Furthermore, we found that NP-based films can act as a medium with a tunable refractive index of up to $\Delta n \approx 0.65$ (at a wavelength of 1.24 μm and energy of 1 eV, respectively).

See supplementary material for detailed information on sample fabrication, NP size distribution, discussion of crystallite size and film roughness of GeTe NP films, effective medium approaches to describe the NP-based films, optical band gap and refractive index change for different PCMs, and additional data on sample aging.

This project was funded by the European Research Council under the European Union's Seventh Framework Program (FP/2007-2013) / ERC Grant Agreement Nr. 339905 (QuaDoPS Advanced Grant). A.U.M. and M.Y. acknowledge support from the Marie Curie ETH Zurich Postdoctoral Fellowship and an Ambizione Fellowship (No. 161249), respectively. The authors thank A. Olziersky and R. Brechbühler for technical assistance and M. Aellen for fruitful discussions. TEM measurements were performed at the Scientific Center for Optical and Electron Microscopy (ScopeM) at ETH Zurich.

**Figures**

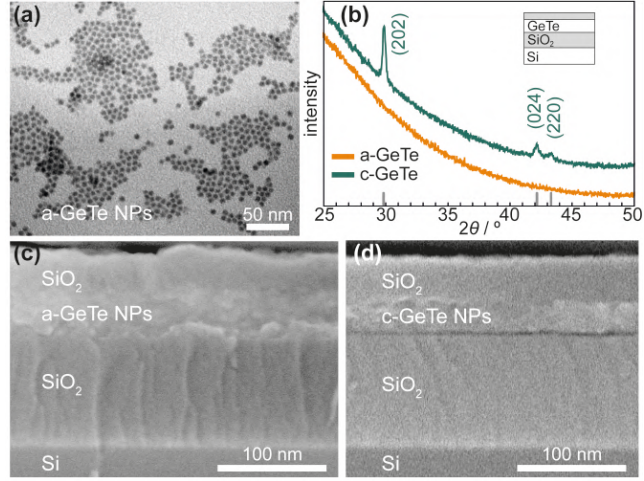

FIG. 1. (a) TEM image of as-synthesized a-GeTe NPs with an average diameter of 5.8 nm. The a-GeTe and c-GeTe thin film samples were characterized via X-ray diffraction in (b) with a scheme of the sample stack in the inset. The peaks are indexed to the rhombohedrally distorted α-phase of c-GeTe (grey ticks according to JCPDS #47-1079). Cross-sections of these samples were further characterized via SEM shown in (c) and (d).

| material | averaged $t_{exp}$ / nm | $t_{fit}$ / nm | material | averaged $t_{exp}$ / nm | $t_{fit}$ / nm |
|---|---|---|---|---|---|
| **sample $S_{a1}$ – as in Figs. 1 to 3** | | | **sample $S_{a2}$ – as in Figs. 2 and 5** | | |
| top $SiO_2$ | 36.6 ± 2.4 | 33 | top $SiO_2$ | 40.2 ± 1.8 | 34 |
| a-GeTe NP film | 41.8 ± 2.4 | 42 | a-GeTe NP film | 35.8 ± 2.6 | 41 |
| bottom $SiO_2$ | 99.8 ± 0.9 | 100 | bottom $SiO_2$ | 101.0 ± 1.3 | 101 |
| **sample $S_{c1}$ – as in Figs. 1 to 3** | | | **sample $S_{c2}$ – as in Figs. 2 and 5** | | |
| top $SiO_2$ | 39.2 ± 1.8 | 33 | top $SiO_2$ | 38.3 ± 2.5 | 34 |
| c-GeTe NP film | 28.9 ± 2.7 | 30 | c-GeTe NP film | 19.7 ± 1.7 | 24 |
| bottom $SiO_2$ | 100.2 ± 0.6 | 101 | bottom $SiO_2$ | 100.8 ± 1.2 | 101 |

TABLE 1. Comparison of the film thicknesses, extracted from cross-sectional SEM images, $t_{exp}$. The averaged $t_{exp}$ (up to 15 values per layer and sample, standard deviations are given with each arithmetic mean) is used as a starting value for the fits. Subsequent fit optimization of the ellipsometry data led to $t_{fit}$, which are in the range of $t_{exp}$, considering the measurement uncertainty of the SEM and the standard deviations for $t_{exp}$.

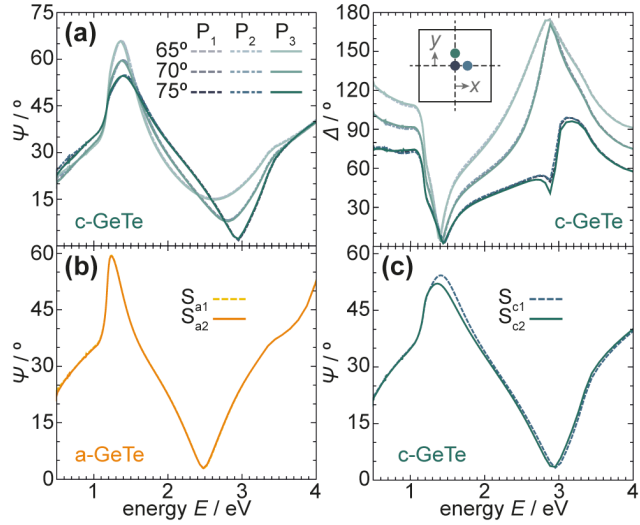

FIG. 2. (a) Spectroscopic ellipsometry results at three angles and at three different sample positions ($P_1$, $P_2$, $P_3$) for the c-GeTe NP sample $S_{c2}$ collected with a VASE® ellipsometer by J.A. Woollam. The probed positions on the sample are marked by colored dots (matching line colors) in the sample scheme displayed as inset in the $\Delta(E)$ data. Ellipsometry results $\Psi(E)$ at 75° are compared for (b) two different a-GeTe NP samples, $S_{a1}$ and $S_{a2}$, and (c) c-GeTe NP samples $S_{c1}$ and $S_{c2}$. On each sample studied for (b) and (c), a position as close as possible to the center position $P_1$ was probed (individual sample mounting might have led to small deviations).

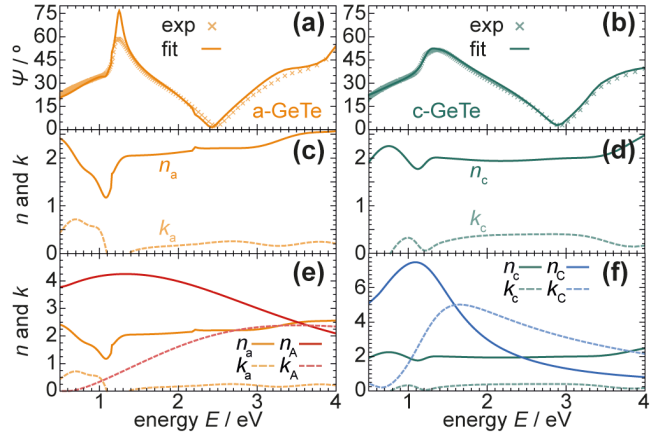

FIG. 3. Spectroscopic ellipsometry data ("exp.", crosses) and fit (line) for $\Psi(E)$ at 75° for the a-GeTe sample $S_{a1}$, orange in (a), and the c-GeTe NP sample $S_{c1}$, green in (b). The extracted refractive indices $n_{a/c}$ (solid) and extinction coefficients $k_{a/c}$ (dashed) are given in (c) and (d), respectively. The data on the NP-based films are compared to literature data on sputtered films for the amorphous – $n_A$, $k_A$ in (e), red – and the crystalline state – $n_C$, $k_C$ in (f), blue.[27]

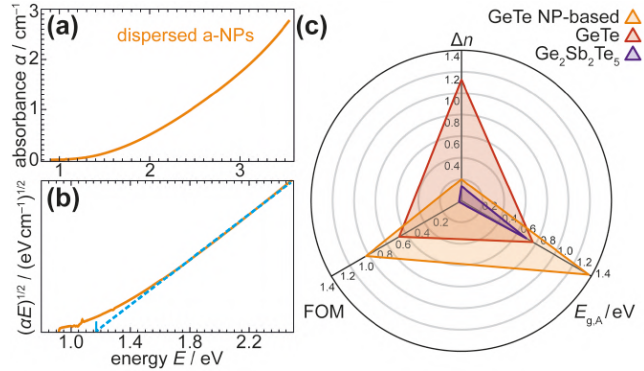

FIG. 4. (a) Absorption spectroscopy of as-synthesized a-GeTe NPs in liquid dispersion. (b) Estimate of the optical (indirect) bandgap $E_{g,a} \approx 1.17$ eV (blue tick) by linear extrapolation (dashed blue) of the absorbance shown in (a). The small feature visible at $E \approx 1.09$ eV in (b) can be ascribed to the organic moiety from the solvent. (c) Radar chart showing the refractive index change $\Delta n$ and the figure of merit $\Delta n/k_A$ at a wavelength of 600 nm, as well as the averaged band gap of the GeTe NP-based film and reference data for sputtered GeTe and $Ge_2Sb_2Te_5$.[51] The axes are arranged such that desirable properties are plotted at larger radii (numbers listed in supplementary material).

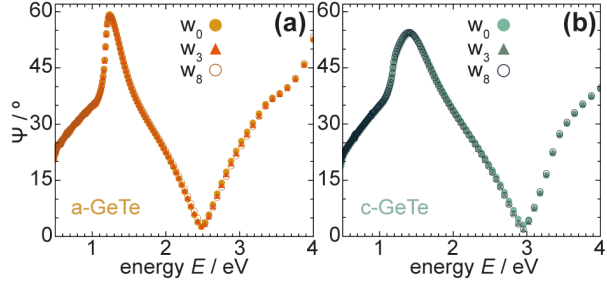

FIG. 5. Spectroscopic ellipsometry results for $\Psi(E)$ at 75° incident angle are compared for (a) the a-GeTe sample $S_{a2}$ (orange) and (b) the c-GeTe NP film $S_{c2}$ (green) after three different storage times at room temperature under nitrogen (one day, three weeks, and eight weeks after preparation – $w_0$, $w_3$, $w_8$). The spectra are identical and no effect of aging on the optical properties was observed.